# Design and Characterization of Self-lubricating Refractory High Entropy Alloy based Multilayered Films

Dawei Luo, Qing Zhou*, Wenting Ye, Yue Ren, Christian Greiner, Yixuan He and Haifeng Wang*

[1] State Key Laboratory of Solidification Processing, Center of Advanced Lubrication and Seal Materials, Northwestern Polytechnical University, Xi'an, Shaanxi 710072, P.R. China

[2] Institute for Applied Materials (IAM), Karlsruhe Institute of Technology (KIT), 76131 Karlsruhe, Germany

[3] Collaborative Innovation Center of NPU, Shanghai 201108, P.R. China

**ABSTRACT**

Refractory high entropy alloys (RHEA) have been proven to have excellent mechanical properties with a potential use as protective thin films. However, the combination of high hardness with low friction and wear is a major challenge in the design of RHEA films. In this study, we show that NbMoWTa/Ag self-lubricating multilayered films give a remarkable reduction in friction and at same time maintain high hardness. Interestingly, it's found that the bcc superlattice dominates in both NbMoWTa and Ag layers and the interfaces become coherent when the individual layer thickness $h$ is reduced below 10 nm. The film properties are then strongly dependent on $h$ ranging from 100 to 2.5 nm, and the most promising properties are obtained when the interface structure transforms from incoherent to coherent one. Specially, the multilayer with $h$ = 2.5 nm exhibits superior tribological performance over monolithic NbMoWTa, due to the significant coherent strengthening along with the self-lubricating ability in the multilayer. This tailored phase transition and coherent structure between the matrix and lubrication phases can also provide an optimal wear rate-coefficient of friction (COF) combination, which is higher than most of the Ag containing self-lubricating films. The current work might open a new route towards the development of innovative self-lubricating RHEA films with excellent tribological properties.

## 1. INTRODUCTION

Since Yeh et al.'s seminal contribution, the concept of alloying several principal elements has been introduced as high entropy alloys (HEAs), which has become a new research hotspot in the past decade due to their unique structures and a multitude of outstanding properties[1,2]. Instead of a mixing of several intermetallics, the alloy is usually observed as a random solid solution with a simple cubic structure[3]. Inspired by the design concept of HEAs and the desire to develop high strength and thermal resistance structural alloys, refractory high entropy alloys (RHEAs) were developed consisting refractory transition metals (four to six groups in the periodic table of elements)[4,5]. Especially, RHEAs like NbMoWTa and NbMoWTaV have been shown to exhibit impressive strength up to 1600 °C[6].

Recently, research interests have begun to focus on the application of HEA thin films, which are a promising way to incorporate RHEAs in harsh environments[7,8]. Due to the limited alloy systems, traditional binary or ternary alloy films in use might not meet the high requirements of modern industry. Exhibiting superior properties over their bulk alloys, HEA films thus show far-reaching prospects towards mechanical, wear and thermal resistance applications[9,10]. Fritz et al. conducted a research regarding the microstructure and mechanical properties of HfNbTiVZr RHEA film at different deposition temperatures[10]. As the deposition temperature rises, the HfNbTiVZr film changes from amorphous to a mixture of Laves and bcc crystalline phases, and its hardness increases from ~6.5 to 9.2 GPa. Sajid et al. reported that the extraordinarily high compression strength of CuMoTaWV RHEA film at room temperature can reach ~10.7 GPa, which constitutes one of the hardest alloy[11]. This high strength is associated with grain refinement effect in the highly textured films with the grain size down to the nanoscale regime[7].

For RHEA films with superior strength and hardness, it's anticipated to have good wear resistance. Unfortunately, the brittle characteristics for most RHEAs always results in crack nucleation during deformation[11,12]. The accumulation of these cracks can cause delamination and material removal, which results in a poor wear resistance. This means that designing of hard yet tough RHEA film is the key to improve their mechanical and tribological properties. On the other hand, lowering the COF is pivotal for RHEAs to improve the stability and efficiency of moving assemblies, but reaching this goal is still technologically challenging[13,14]. For RHEA films to be widely used, pathways to lower the friction forces must be provided.

To improve the friction and wear properties of HEAs, designing self-lubricating composites looks like to be an effective way. With the introduction of a soft second phase into the HEA matrix, the composites provide effective solid lubrication during the long-term operation[15,16]. However, an evident disadvantage inevitably appears, e.g., the incorporation of solid lubrication phases will certainly degrade the high strength of the composites[17]. One efficient approach to solve this problem is to design a bio-inspired laminated composite, as such a layered architecture provides superior resistance to deformation when compared with their constituents[18-20]. In addition, the mechanical and tribological properties can be elaborately modified by tuning the layer thicknesses of the two constituents[21]. In this sense, such a structural design strategy can provide a direction for the development of hard yet tough RHEA self-lubricating films. A prior knowledge of the corresponding deformation mechanisms is important for the design of these wear-resistance composites.

To this end, we here reveal that in a prototypical RHEA NbMoTaW film, the introduction of Ag layers can dramatically enhance the tribological properties of the nano-multilayers, as Ag is known to serve as an effective lubricant phase in reducing friction and wear[22]. More importantly, the lattice parameter of close-packed planes in Ag is quite close to that of NbMoTaW, which is promising for preserving the high hardness by carefully design coherent multilayers. Furthermore, this paper attempts to correlate the layer thickness with the mechanical as well as tribological behavior of these novel NbMoTaW/Ag multilayers, and an in-depth understanding of the potential deformation and wear mechanisms is also provided.

## 2. EXPERIMENTAL

The alloy target of NbMoTaW with an equimolar composition was fabricated via a powder metallurgy technique, as shown in Figure 1. A powder mixture of pure elements with an nominal purity of 99.9 wt.% was ball milled with $Si_3N_4$ balls for an hour. Then, under an argon atmosphere, the alloyed powder was consolidated by spark plasma sintering (SPS). On the other hand, the Ag target with purity better than 99.9 wt.% was obtained by arc melting in Ar atmosphere. After that, at ambient temperature, the NbMoTaW/Ag multilayers containing equal sublayer thicknesses (i.e., the thickness of NbMoWTa layer is same as the Ag layer) were deposited on Si substrates. The thicknesses $h$ are chosen as 2.5, 5, 10, 20, 50 and 100 nm for different multilayers. Sputtering chamber background pressure was $6.3\times10^{-5}$ Pa and the operating pressure was kept at $5.4\times10^{-1}$ Pa (argon gas flow rate of 3 sccm) during sputtering. To enhance the homogeneity of the film, the substrate holder was rotated during deposition and a bias voltage of −80 V was applied to the substrate during all depositions. The total thickness of each of multilayers was controlled to be approximately 1.6 μm and the top layer was NbMoTaW. For comparison, pure NbMoTaW and Ag films were also deposited at same conditions. Prior to deposition, the silicon wafers were rinsed in an ultrasonicator (ethanol environment) for 20 min, and then air-dried.

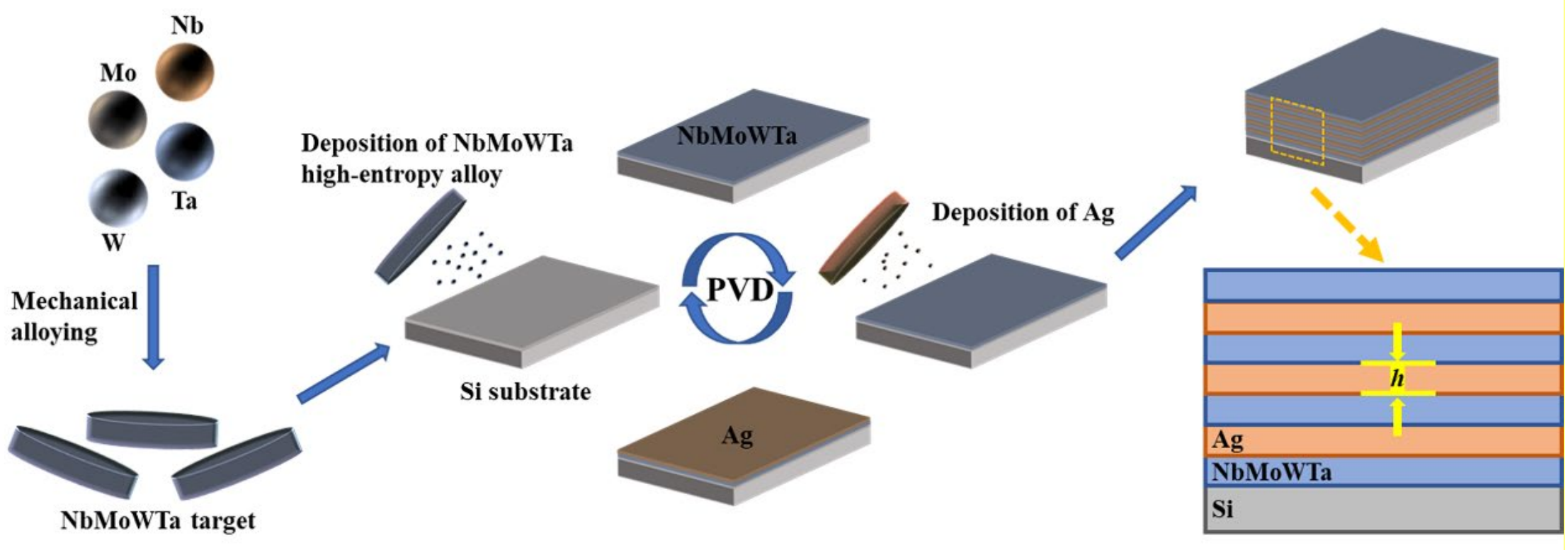

**Figure 1.** Schematic diagram illustrating the fabrication of the multilayer film with alternating layers of NbMoTaW RHEA and Ag.

The crystallographic texture of each specimen was analyzed by X-ray diffraction (XRD Bruker D8 with Cu $K_\alpha$ radiation) with a scan rate of 0.02° $s^{-1}$. The modulation cross-sectional morphology of the multilayers was acquired with scanning electron microscope (SEM; ZEISS GeminiSEM 500) and transmission electron microscopy (TEM, Talos F200X G2). Besides, the chemical compositions were analyzed by energy-dispersive X-ray spectroscopy (EDX) in SEM and TEM. For the TEM analysis, the specimen was cut into thin discs of 3 mm in diameter, manually polished and finally thinned by ion-milling using Ar ions. AFM measurements (Oxford MFP-3D) were performed in a tapping mode using a monocrystalline silicon probe with a nominal tip radius of ~ 10 nm.

The mechanical behavior was examined by nanoindentation (Nanoindenter XP system) with a Berkovich tip[23,24]. After calibrated with standard fused silica, the tip radius of the Berkovich probe was estimated roughly as 80 nm. The modulus $E$ and hardness $H$ of the samples were obtained by a depth control mode under a continuous stiffness measurement. The preset depth was 200 nm (one eighth of the total film thickness) to make sure that at least one interface was involved and there was no substrate effect during the indentation process[25]. The intrinsic $E$ and $H$ were calculated from each indentation depth profile, where a plateau was reached[26]. On each sample, averaged values based on 16 indentations were used.

Tribological tests were performed in a dry environment using a Ball-on-Block apparatus (Rtec

Corporation Ltd.)[27]. The samples were sliding against a $Si_3N_4$ counterbody (6.25 mm in diameter) in a circular path under a normal load of 0.5 N. The circular radius, rotation speed and sliding time were fixed as 1.5 mm, 20 r $min^{-1}$ and 20 min, respectively. The tribological tests of all monolithic NbMoWTa film and NbMoWTa/Ag multilayers were repeated at least four times to ensure reproducibility. The surface profiler (NPFLEX, Bruker) with white light interference was used to scan the wear tracks and calculate the volume loss. In order to get statistical results, each measurement was repeated at least four times per sample. After the dry sliding wear tests, the wear track of each sample was examined by SEM to reveal the wear mechanism. The compositions at the worn surface were analyzed by X-ray photoelectron spectroscopy (XPS) with a standard Al $K_\alpha$ source (PHI 5000 VersaProbe). The lubricating films were studied by high resolution spectra obtained by 1 keV Ar+ ions etching for 10 min to remove surface contaminations. The spectrometer was calibrated by setting the binding energy of the hydrocarbon C 1s line to 284.8 eV with respect to the Fermi level.

## 3. RESULTS AND DISCUSSION

### *3.1 Microstructural characterization*

In Figure 2a, SEM analysis of the NbMoWTa film on the Si substrate shows a needle-like morphology, being in agreement with the general observation in other RHEA films prepared by magnetron sputtering[8,11]. The cross-sectional morphology of NbMoWTa shows a dense and columnar structure, as demonstrated in Figure 2b. Besides, a thin amorphous layer appears at the interface between the Si substrate and the deposition film, which was also detected in other RHEA films[11]. The interfacial stress between the film and the substrate is considered as the reason for the appearance of this amorphous transition layer[11], caused by the huge lattice mismatch (about 41%) between Si (lattice constant of 5.43 Å) and NbMoTaW (lattice constant of 3.24 Å). The EDS mapping of the film surface (Figure 2c) shows a homogeneous distribution of all the related elements with a nearly equiatomic composition, indicating no segregation or precipitation of a second phase. Table 1 shows that the element contents of NbMoWTa are almost equal. AFM analysis (Figure 2d) showed that the thin film formed elongated nanocrystalline grains[11] and the average surface roughness (RMS) was ~ 3 nm.

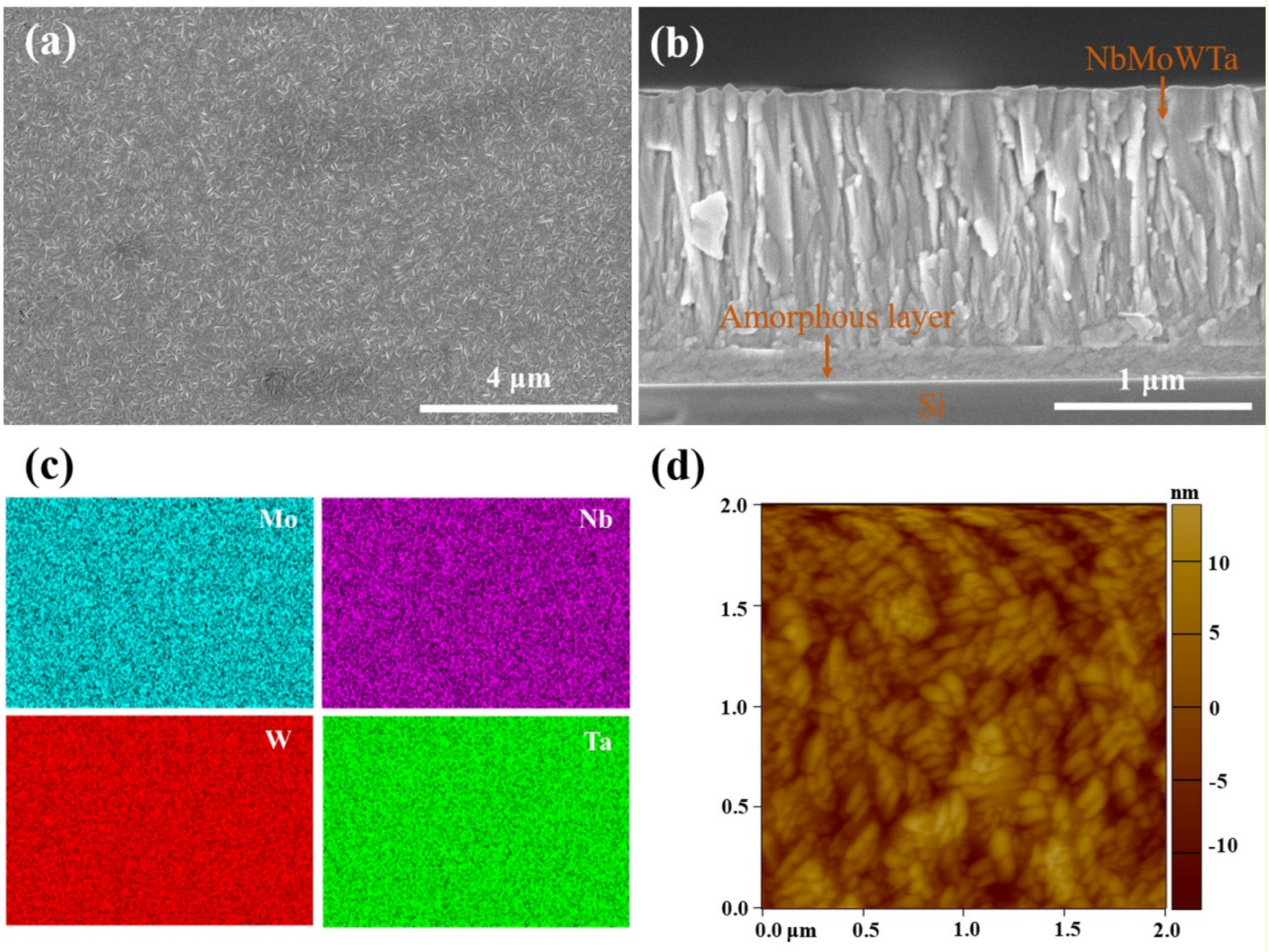

**Figure 2.** Microstructure of the NbMoTaW RHEA film: (a) SEM image of the deposited surface, (b) cross-section morphology, (c) element distribution mapping of the film surface and (d) AFM image of the film surface.

**Table 1. EDS area analysis of the NbMoWTa film (at. %)**

| Element | Nb | Mo | W | Ta |
|---|---|---|---|---|
| Content | 23.2±0.1 | 24.3±0.2 | 26.2±0.2 | 26.3±0.1 |

The XRD diffractograms of all the films are shown in Figure 3. The NbMoWTa RHEA film shows that a single bcc solid solution phase is formed with a (110) preferred orientation, and the Ag film has a strong (111) texture. Further, the XRD analysis of the multilayer with $h$ = 100 nm shows no change of fcc Ag (111) and bcc NbMoWTa (110) textures and no generation of additional phases. As $h$ is further decreased to 5 nm and below, the Ag (111) and NbMoWTa (110) peaks decrease and nearly disappear. At the same time, an extra diffraction peak with a higher intensity appears in-between the (111) and (110) peaks, indicating that the crystallinity of NbMoWTa /Ag multilayers are better than that at $h$ = 10 nm, and the coherent superlattice are formed[25].

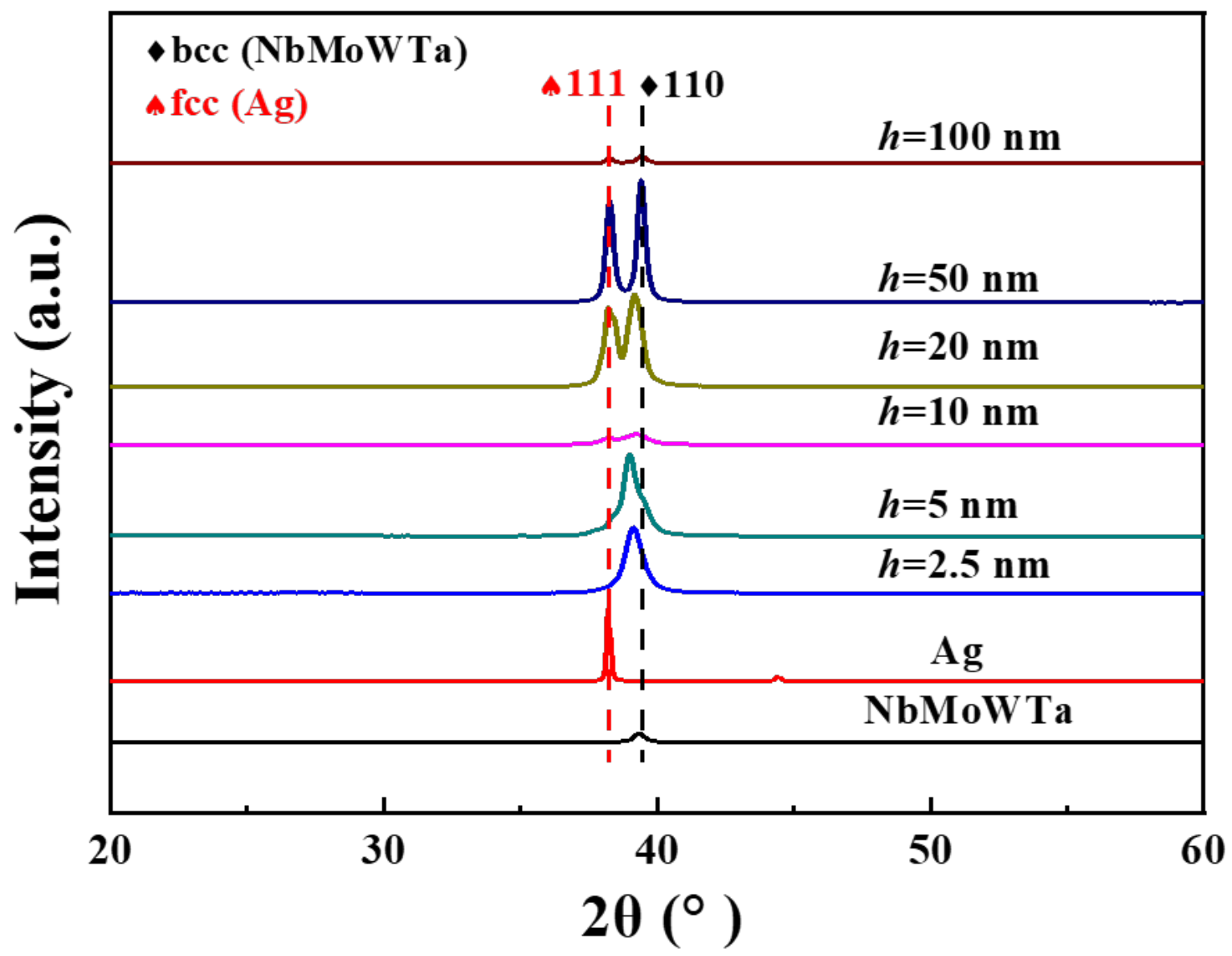

**Figure 3.** XRD diffractograms of the NbMoWTa, Ag monolithic films and their multilayers

Cross-sectional SEM images (Figure 4) of the NbMoWTa/Ag multilayers show distinct microstructural features at different length scales. Compared with the monolithic films, the columnar structure of NbMoWTa is broken in the multilayers with a large *h* due to the presence of Ag layer. Instead, a clear layered structure is presented in the Figure 4a,b. As the individual layer thickness of the NbMoWTa/Ag multilayer decreases to 5 nm (Figure 4c), the columnar structure extending throughout the whole film appears again. When *h* is small enough, the heterogeneous layers would become coherent, which facilitates the formation of a superlattice columnar structure[25]. The corresponding schematic illustration is presented in Figure 4d. In conclusion, we observe two types of microstructures in the NbMoWTa/Ag multilayers, which correspond to different layer thickness ranges, i.e., when $h > 5$ nm the multilayer presents a layered structure, otherwise it forms the coherent columnar structure.

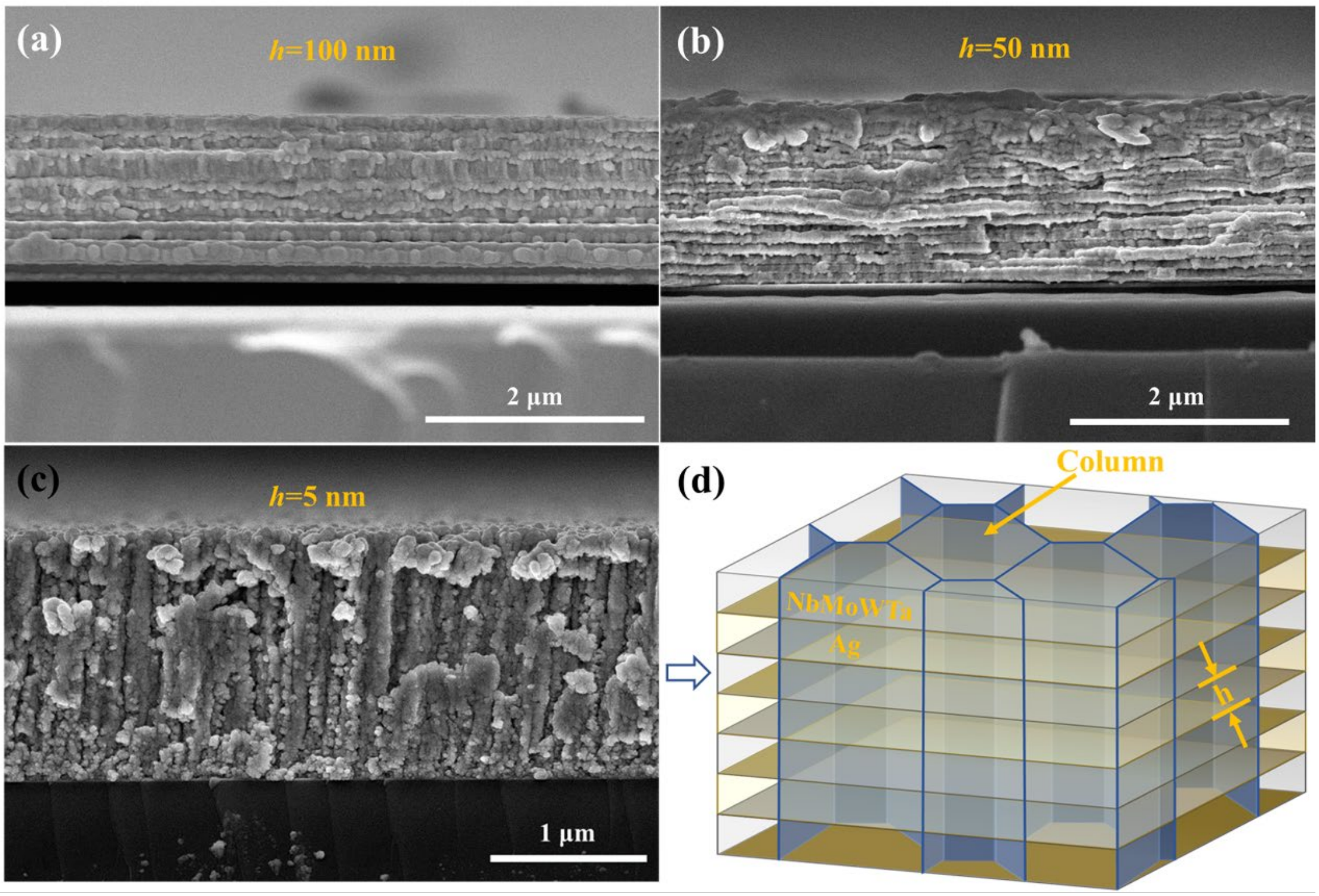

**Figure 4.** Cross-sectional SEM images of the NbMoWTa/Ag multilayers: (a) $h$ = 100 nm, (b) $h$ = 50 nm and (c) $h$ = 5 nm. (d) The schematic diagram showing the formation of the coherent columnar multilayer.

In order to get more detailed structure information of the multilayers, TEM analyses were further carried out. High-angle annular dark field-scanning transmission electron microscopy (HAADF-STEM) and the corresponding EDS mapping of all the elements in the multilayer with $h$ = 100 nm are shown in Figure S1 (Supplementary), which displays obvious contrast difference between the alternative NbMoWTa and Ag layers. Clear interfaces between both constituent layers can be observed, consistent with the aforementioned SEM analysis. The cross-sectional TEM image of the multilayer with $h$ = 20 nm is shown in Figure 5a, in which both layers with a well-modulated structure are identified. The corresponding selected area diffraction patterns (SAED) show diffraction rings, indicating the polycrystalline nature of Ag and NbMoWTa layers. The dark-field image obtained from the diffraction spot along the film growth direction is further shown in Figure 5c. It indicates that the nanometer grains inside the constituent layers are restrained by the layer interface, leading to nearly equiaxed grains rather than columnar grains generally observed in monolithic films[28,29]. Figure 5d shows HRTEM analysis of the multilayer and it can be seen that the heterogeneous interfaces are dense and chemically sharp. Besides, a wide amorphous grain boundary formed in NbMoWTa is also seen. EDS analysis at this region shows a high content of Ag and significant intermixing of both layers during the nonequilibrium deposition process. As Ag atoms are easier to diffuse into the grain boundaries in the NbMoWTa layer, an amorphous structure is therefore developed at the region where five kinds of elements are mixed[30,31]. Finally, the Fast Fourier Transforms (FFTs) of the framed regions in Figure 5d are presented in Figure 5e,f. The diffraction zones of [011] and [-111] corroborate that Ag and NbMoWTa possess a stable fcc and bcc structure with a strict K-S orientation relationship of {111} // {110}.

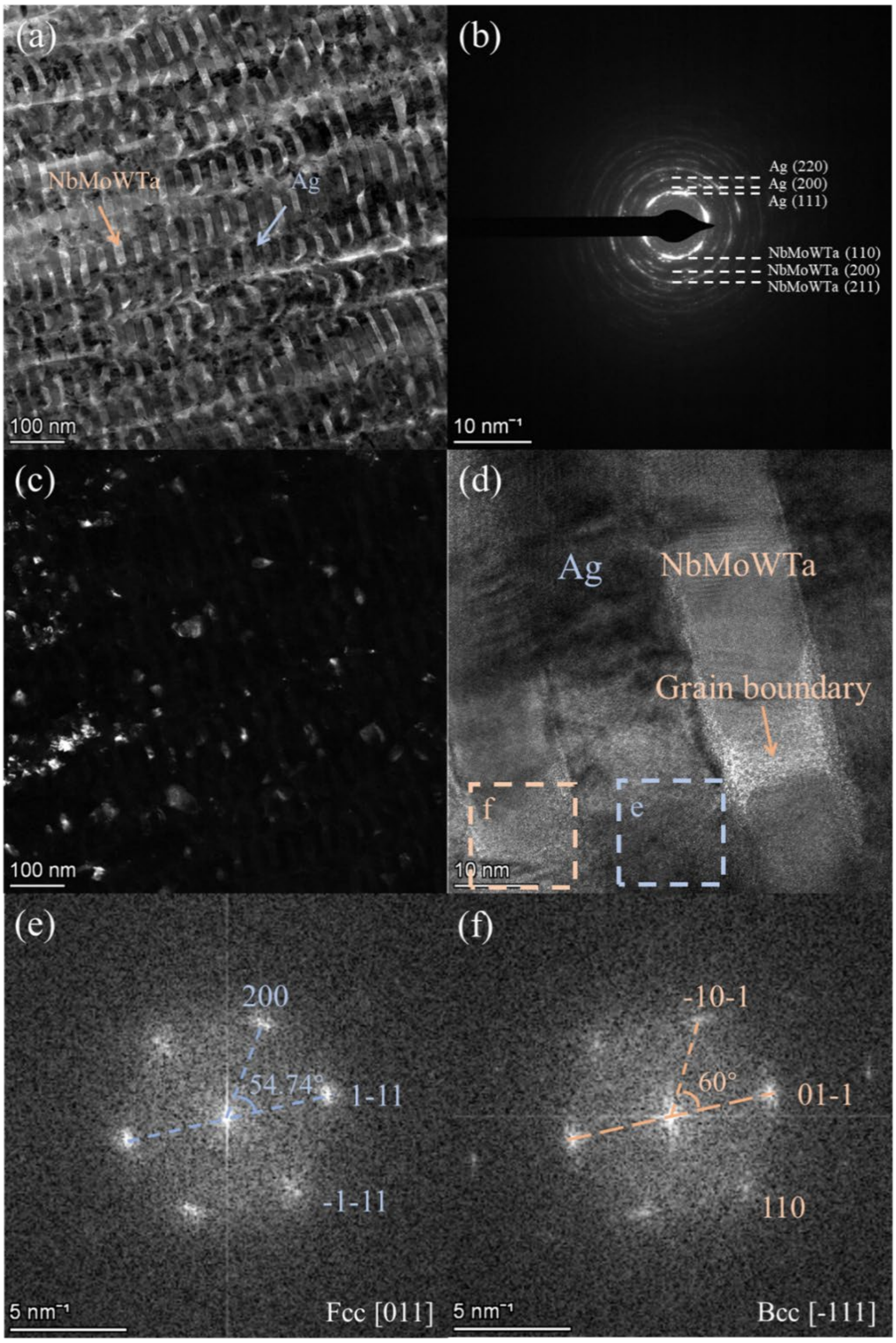

**Figure 5.** Representative cross-sectional TEM images of NbMoWTa/Ag multilayer (*h* = 20 nm) including (a) Bright-field, (b) SAED pattern, (c) dark-field and (d) high resolution TEM images. The boxed regions in (e) and (f) correspond to the FFT patterns of NbMoWTa and Ag layers, respectively. The orange arrows indicate the grain boundary region formed inside the constituent layer

HAADF-STEM and corresponding EDS mapping of the NbMoWTa/Ag multilayer (*h* = 5 nm) are shown in Figure S2. It is found that the chemically distinct layered structure can still be maintained at such a small scale. The corresponding bright field cross-sectional TEM images are presented in Figure 6a. By means of the epitaxial growth of Ag on the NbMoWTa layer, the superlattice columnar structure is formed. The interfaces inside each columnar grain are coherent, forming a quasi-single crystal structure[32]. Besides, the position outlined by the orange dotted line reveals that the layered structure breaks at the columnar grain boundary (CGB). The statistically obtained column size follows a normal distribution with a size range of 120 nm ~ 160 nm, which is nearly 30 times of the layer thickness. Similar results were also reported in Cu/Ru multilayers where coherent interfaces are formed[25]. Figure 6b shows the SAED of the multilayer with *h* = 5 nm. Different from the diffraction rings in the case of *h* = 20 nm (Figure 5b), the diffraction spots are sharp along the growth direction of the multilayered film (outlined by white circles), which indicates that the samples with small *h* have a preferential orientation[32]. The dark-field image shown in Figure 6c also indicates that the columnar grain can run through each constituent layer (indicated by dotted line), which is in strong contrast with that shown in Figure 5c. Figure 6d and Figure S3 present the HRTEM images of the heterogeneous interfaces between NbMoWTa and Ag. The FFT and inverse FFT (IFFT) patterns acquired from the yellow area are shown in Figure 6e,f, respectively. These results denote that a bcc stacking sequence appears continuously across both constituent layers inside the column grain, and coherent interfaces are formed when $h \leq 5$ nm.

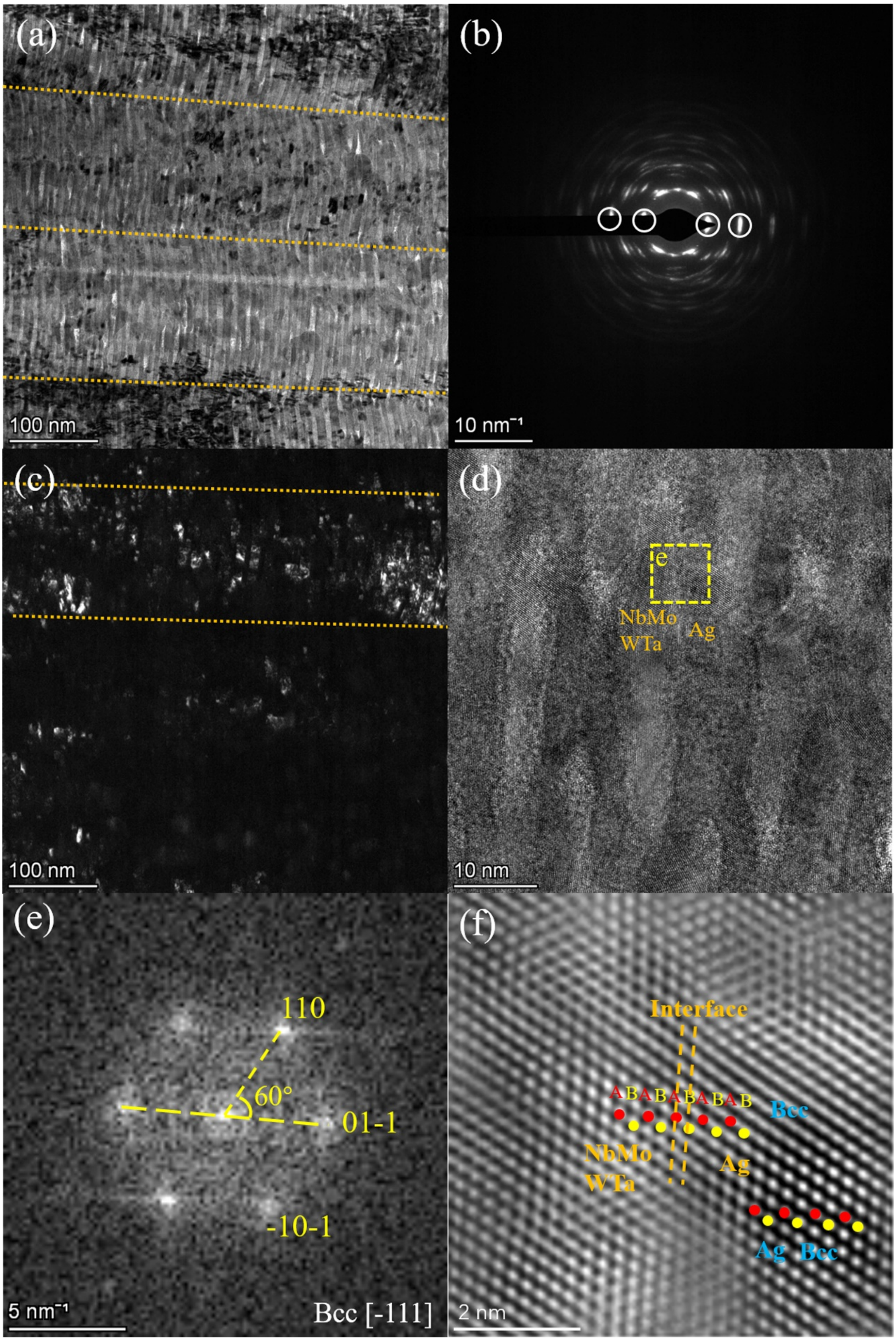

**Figure 6.** Typical cross-sectional TEM images of NbMoWTa/Ag multilayer with $h$ = 5 nm including (a) bright-field, (b) selected area electron diffraction pattern, (c) dark-field and (d) high resolution TEM images. The FFT (e) and IFFT (f) patterns are corresponding to the selected area marked by the yellow frame in (d). Dotted orange lines in (f)

show the location of the heterogeneous interface.

Generally, as the size of the material is reduced to a nanometer scale, metastable crystalline structure is formed. For metallic multilayers, when $h$ drops below a critical value, the transformation of lattice structures often takes place to accommodate lattice mismatch of two constituent layers [25,33,34]. For the NbMoWTa (110) and Ag (111) planes, the lattice mismatch is only ~ 2.8%. Such a small mismatch at a heterogeneous interface will give rise to a high tendency for the construction of coherent multilayers[25,34,35]. Accordingly, two types of interfaces, i.e., incoherent Ag-fcc/NbMoWTa-bcc and coherent Ag-bcc/NbMoWTa-bcc interfaces, are identified when the layer thickness decreases toward the nanoscale.

The phase transitions of multilayers could be explained by a thermodynamic model[35,36]. Considering the total effect of increase in volumetric energy and the reduction in interfacial free energy as a result of phase transition, the model proposes that the total energy of a unit bilayer system should be lowest. When the structure of Ag undergoes a transition to bcc, the total normalized free energy change $\Delta g$ could be expressed as follows:

$$\Delta g = 2\Delta\gamma + \Delta G_{Ag} h \qquad (1)$$

Here $\Delta\gamma$ is the change in the interfacial energy between incoherent and coherent interfaces, and $\Delta G$ represents the allotropic free energy change of the metastable Ag. Using the experimental results obtained in this study, as $h$ is large, $\Delta g$ should be 0 if the Ag layers maintain their stable structure. When $h$ decreases to or below 5 nm, the Ag lattice begins to transform from the bulk-stable structure to a meta-stable structure due to the "template effect"[37,38] of the underlying NbMoWTa layer. Driven by the tendency to form coherent interfaces with low interfacial energy ($\Delta\gamma < 0$), the transition of Ag from fcc to bcc is due to the fact that the energy of the system can be minimized when the increase of volumetric free energy is lower than the reduction of interfacial free energy (i.e., $h_{Ag} < -2\Delta\gamma/\Delta G_{Ag}$). This provides a sound explanation for the observed lattice structure transition when $h$ is reduced in the NbMoWTa/Ag multilayers. Apart from this thermodynamic consideration, the stacking faults (SFs) formed (Figure 6a) and the stress created by the misfit strain at the interface also promote this structural transformation of Ag[39,40].

Other than the phase transformations aforementioned, with the decrease of $h$, the degree of texture, epitaxy and coherency of the NbMoWTa/Ag are found to increase. For multilayers, there exists a critical layer thickness, below which arrays of misfit dislocations are absent and the interface can be coherent. This critical sublayer thickness is approximately calculated as[25,41]:

$$h_c = b\left(In\left(h_c/b\right)+1\right)/\left(8\pi\varepsilon\left(1+\nu\right)\right) \tag{2}$$

Here $h_c$ is the critical sublayer thickness, $b$ represents the Burgers vector, $\varepsilon$ is the lattice mismatch, and $v$ is the Poisson ratio. Hence, $b$ = 0.29 nm[19], $\varepsilon$ =0.028 and $v$ = 0.3 are chosen to estimate $h_c$ in Eq. (2). The value is then calculated to be ~ 1 nm, which is much lower than 5 nm derived from the experiments. The theoretical values typically underestimate the experimental ones as was also reported in the literatures[25,41]. This discrepancy might be caused by factors such as the stacking faults formed in Ag, the elastic isotropy and linear elastic behavior in the lattices[41,42]. These factors can alleviate the large amount of elastic strain energy stored at the interfaces. Summarizing, it is evident that adjusting the sub-layer thickness is an effective way to control the phase and interface structures of NbMoWTa/Ag multilayers.

*3.2 Mechanical properties*

Mechanical characterization of the monolithic NbMoWTa film and multilayers is conducted through nanoindentation testing. The nanocrystalline nature of NbMoWTa results in a high hardness of ~ 10.9 GPa, which is much higher than the hardness of the respective bulk alloy, but is comparable with the reported literature values for RHEA films[43,44]. It's believed that the superposition effects of solid solution hardening and column boundary strengthening presumably contribute to this measured high hardness in NbMoWTa films[43,45,46]. The values of nano-hardness $H$ for different multilayers are shown in Figure 7a, according to which the hardness is found to increase with decreasing $h$. At $h$ < 10 nm, the hardness of multilayers shows a sharply rising trend, with the maximum value 52% higher than that calculated following the rule of mixtures (ROM) of the NbMoWTa and Ag films. Figure 7b shows the elastic modulus of NbMoWTa/Ag multilayers. According to the elastic moduli of single phase Ag film ($E$ =82 GPa) and NbMoWTa film ($E$ =214 GPa), the rule of mixture results in $E_{ROM}$ =148 GPa, which is nearly the same as the NbMoWTa/Ag multilayer with $h$ = 100 nm. Interestingly, the modulus values decrease as $h$ is reduced, but a sharp increase occurs when $h$ < 10 nm, leading to values larger than $E_{ROM}$.

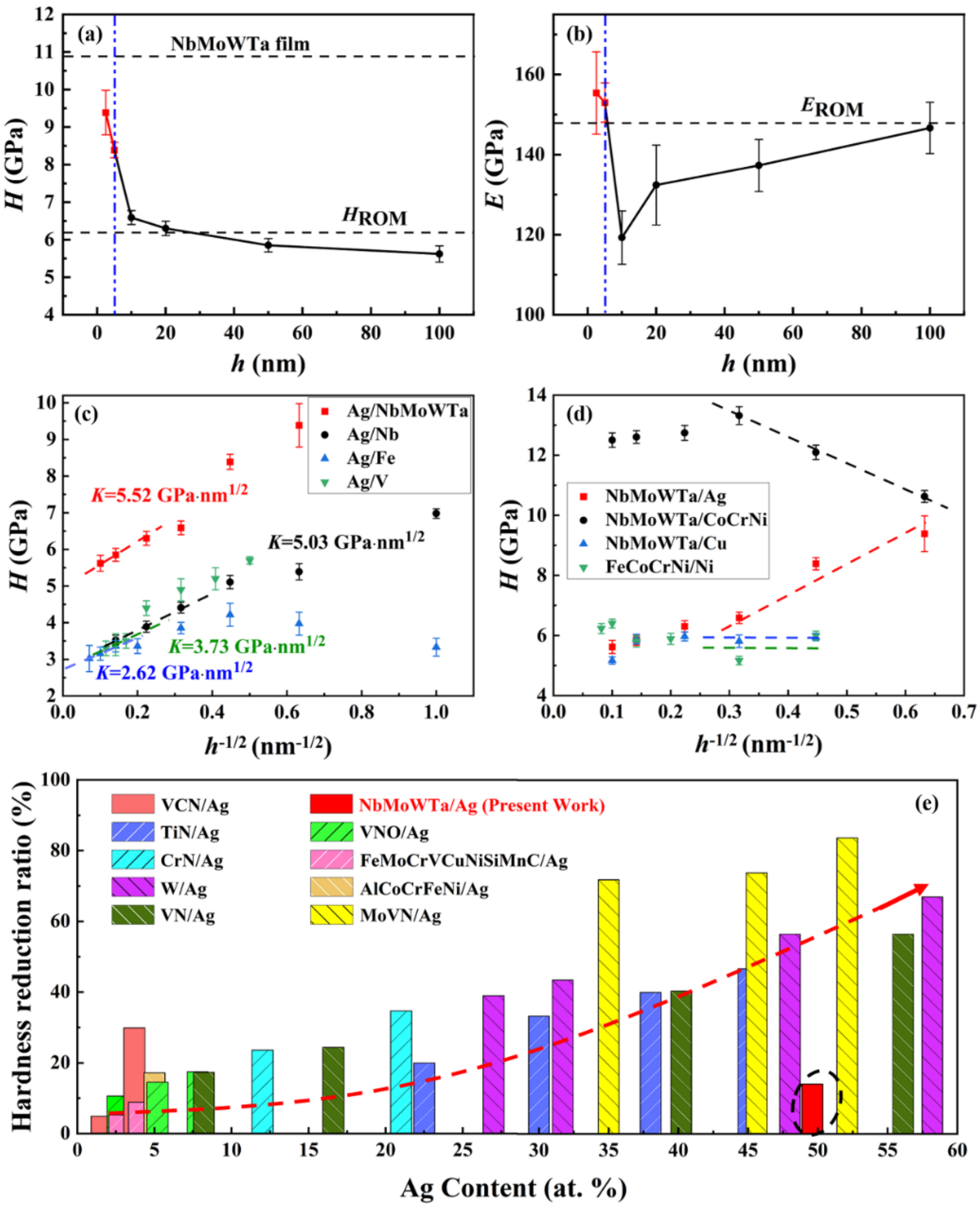

**Figure 7.** The dependence of indentation hardness (a) and elastic modulus (b) on $h$ for Ag/NbMoWTa multilayers. (c) According to literature reports, the hardness of Ag/X (fcc/bcc bimetallic multilayers) varied with $h^{-1/2}$ and its Hall-Petch slop are shown. (d) Hardness of HEA/X bimetallic multilayers varied with $h^{-1/2}$, with special attention on the variation trend of $H$ at small scales (e.g. $h < 20$ nm). (e) Hardness reduction ratio of Ag containing materials reported in literature as a function of Ag content. Refs of other Ag based fcc/bcc bimetallic multilayers[47-49], HEA based bimetallic multilayers [31,50,51] and Ag contained self-lubricating films [52-60].

Before presenting the underlying mechanisms responsible for the noticeable layer thickness dependent mechanical properties, we first focus on the relationship between the hardness of NbMoWTa/Ag multilayers and $h^{-1/2}$ in Figure 7c. When $h$ is large, the hardness increases as a linear function of $h^{-1/2}$, conforming to Hall-Petch law expounded by the pile-up of dislocations at interfaces. This behavior of NbMoWTa/Ag was also found in other Ag based fcc/bcc bimetallic multilayers[47-

[49], as shown in Figure 7c. The calculated $K$ (Hall-Petch slop) of Ag/NbMoWTa is 5.52 GPa·nm$^{1/2}$, which is a little bit larger than those reported for most Ag based fcc/bcc multilayer systems. Besides, the value is also found to be higher than $K$ = 4.2 GPa·nm$^{1/2}$ for the nanocrystalline Ag (with grain size from 17 to 105 nm)[19,61]. These results indicate that in RHEA multilayer the strengthening ability caused by pile up of dislocations at interfaces is relatively higher. During plastic deformation, the two layers deform inhomogeneously, and back stresses in the form of dislocation pile-ups[62] are generated in the soft Ag layers. This back stress can counteract the applied stress to prevent dislocation emission and slip, which makes Ag appear to be stronger. It's reported that large difference in hardness between the two constituent phases can cause a superior pining ability of dislocations by the interface[18,51]. Therefore, when intersecting with the soft metal to build a multilayered composite, the HEA can impose a higher hardening rate than its constituents due to its excellent mechanical performance.

Although the HEA layer can provide a basis for the excellent mechanical properties of the multilayers, the characteristic of soft fcc layer can also have a huge influence on the strengthening mechanisms especially at small length scales. When $h$ is further reduced, the hardness values of NbMoWTa/Cu, NbMoWTa/CoCrNi and CoCrFeNi/Ni either decrease or level off[31,50,51], while a continuous increase in hardness of the NbMoWTa/Ag multilayer can be observed (Figure 7d). Generally, when the individual layer thickness or the grain size decreases below a few nanometers in multilayers[63] or in nanocrystalline metals, the mechanism known as the inverse Hall-Petch effect dominates[64]. Interestingly, by contrast, the present NbMoWTa/Ag multilayers show a continuous increase of $H$ as $h$ is less than 10 nm, which is beyond the expectation in light of conventional understanding in materials science.

Generally, the peak hardness of a metallic multilayer could be estimated from the interface barrier strength $\tau$ as[25]:

$$\tau = (K/M)^2 \pi(1-\nu)/\mu b \qquad (3)$$

Here $M$ is the Taylor constant and $\mu$ is the shear modulus. For NbMoWTa/Ag multilayers, the stress concentration at the interface is caused by dislocation pile-ups in the Ag layer. When this stress concentration exceeds the limit, the transfer of dislocations from Ag to NbMoWTa layer would occur. Hence, $\mu$ = 30 GPa, $K$ = 5.52 GPa·nm$^{1/2}$, $M$ = 3.1, $\nu$ = 0.3 and $b$ = 0.29 nm[19] are chosen to estimate $\tau$ in Eq. (3). The peak hardness of the multilayer is then obtained by multiplying shear stress with the Taylor factor to convert to a strength value and then multiplying the strength by 2.7 to convert to a nanoindentation hardness[26]. The value is then calculated to be 6.7 GPa, which is about 29% lower when compared with 9.39 GPa derived from the experiments. Instead, the measurements for most other fcc/bcc incoherent systems do not differ much from the calculated peak values[26,63,65]. As the structure transition effect is ignored in the derivation of Eq. (3), this discrepancy for the NbMoWTa/Ag multilayer demonstrates that the change of interface structures does indeed affect the mechanical properties and deformation mechanisms.

When $h$ is reduced, there is no enough space for dislocations to pile-up at the interfaces[66,67]. Then discrete dislocation slip starts to dominate plastic deformation, and plastic yielding is reached when gliding dislocation starts to transmit across the interface, spreading plastic deformation to adjacent layer. In this circumstance, the blocking of dislocation motion by hetero-interfaces therefore determines the hardness of the multilayers[26,68]. For the present NbMoWTa/Ag multilayers, a fully coherent superlattice is established at this length scale as both constituents adopt the same crystalline structure (Figure 6). Therefore, in addition to the ROM type hardness, the coherent strain developed at the interface will provide additional hardening to dislocation motion. The theory of coherent strain[19,69,70] predicts that there is significant strengthening developed at coherent interfaces due to the presence of lattice mismatch. When interatomic bonds are stretched or compressed to match the lattice of adjacent layers, a coherent stress filed of alternating tensile/compressive stress are created. In this condition, the slip of dislocations would be heavily obstructed by these alternate stresses at the interfaces, which then accounts for the excellent hardness enhancement in the multilayer. Consequently, as $h$ decreases, the density of interfaces increases and the coherent strengthening will be more significant. For the present NbMoWTa/Ag, this will suppress the softening generally observed in other fcc/bcc multilayers with small $h$.

From above analysis, it can be concluded that the deformation mechanism of NbMoWTa/Ag multilayers change from classic Hall–Petch strengthening to coherent strengthening with a continuous decrease of $h$ (differentiated by the blue dotted line in Figure 7a). The singularity of the hardness at sufficiently small $h$ is mainly dependent on the evolution of the interface structure. On the other hand, according to Eq. (2), a moderate interface misfit is required to achieve coherency in fcc/bcc multilayers as the increase of $\varepsilon$ would lead to a decreased $h_c$. This indicates that if the interface misfit is too large, coherency is hard to be achieved in these systems because the misfit dislocations are so close together that their cores overlap[71]. In general, coherent interfaces are mostly reported in multilayers with lattice mismatch less than 5%[19]. For NbMoWTa/CoCrNi and NbMoWTa/Cu multilayers, the lattice misfit between adjoining metallic layers reach 10% and 9%, respectively. Accordingly, the phase structure transformation and coherent lattice formation are hard to be observed in these two systems, leading to a failure of continuous strengthening with decreasing $h$[31,50].

Due to this special coherent strengthening at small scales, it is noticed that peak hardness of NbMoWTa/Ag can reach a high value with only 14% reduction in $H$ when compared with the monolithic NbMoWTa. As the addition of a lubricating phase will inevitably deteriorate the mechanical properties, the hardness reduction ratio is found to increase greatly with the Ag content in self-lubricating films, as shown in Figure 7e[52-60]. However, by virtue of the unique structural design, the NbMoWTa/Ag multilayer can overcome the tendency of substantial decrease in hardness (outlined by black circles), although a large number of Ag has been added.

Finally, the non-monotonic evolution of $E$ with the layer thickness is discussed as follows. As shown in Figure 7b, $E$ shows a drastic softening of 20% as $h$ is reduced. The reduced modulus of multilayer is considered for two reasons. First, the elastic modulus are generally lower for metallic films with tiny grains because of the large GB density[72]. Likewise, it's believed that for incoherent multilayers, there may exist interfacial layers whose disordered structure can have an obvious inelastic strain as compared to the constituent crystalline layer[18,73]. Accordingly, the compliant incoherent interfaces can affect the modulus of a multilayer in an approximate manner as[63,74]:

$$1/E=2h\,(2h_i/E_i+(2h-2h_i)/E_0)^{-1} \tag{4}$$

here $h_i$ and $E_i$ are the thickness and modulus of interface respectively, and $E_0$ is the ROM value of both constituents. The multilayer will have more interfaces with the decrease of $h$, leading to the decrease of $E$.

Second, as mentioned in Sec. 3.1, the amorphous intermixing regions can be formed with the decrease of $h$[31,75]. It is reported that the specific volume ("free volume") of amorphous is 1% larger than the crystalline state[76], which then imposes a negative effect on the modulus of the system due to a looser atomic packing. As a result, these amorphous can also account for the reduction of $E$ with the decrease of $h$[77-78]. While, as $h$ is smaller than 10 nm, the NbMoWTa/Ag multilayers exhibit a significant increase in $E$, which may be attributed to the compressed interplanar spacing of constituent layers[65,79]. From the XRD results presented in Figure 3, the high-angle shift of the main peak indicates the contraction of lattice spacing, causing modulus enhancement as $h$ is reduced to 2.5 nm. Furthermore, the highly textured coherent structure can also enhance the modulus, as reported previously[49].

*3.3 Tribological properties*

The tribological tests of NbMoWTa film against the steel counterbody (hardness of 600 HV) show an inappreciable wear on the order of $10^{-7}$ $mm^3/Nm$, indicating that the film has superior wear resistance against steel. Consequently, in order to clarify the wear behavior, the test was carried out with a $Si_3N_4$ counterbody (hardness of 1550 HV) to induce significant wear on the film. On the other hand, the wear tests of the multilayer with $h \geq 50$ nm show excess wear and removal of the films from the substrate. Therefore, they are excluded from further investigations. A plot of the coefficient of friction versus the sliding time for all the films is shown in Figure 8a. The multilayers show a lower COF when compared with the monolithic NbMoWTa film. Figure 8 also demonstrates the duplex effect of $h$ on the tribological properties of NbMoWTa/Ag multilayers. It can be seen that upon decreasing $h$, the COF increases but the wear rate decreases significantly spanning nearly two orders of magnitude. Especially, it shows a low average wear rate of $9\times10^{-6}$ $mm^3/N$ m when the individual layer thickness $h$ is 2.5 nm, which is nearly half of that in the monolithic NbMoWTa film. In order to get information on the worn surfaces of the films after the tests, white light interferometry morphologies are given in Figure 8c and Figure S4. The wear tracks of the multilayers are continuous and smooth. However, severe local flaking and delamination occurred in the monolithic NbMoWTa film, resulting in a reduced surface integrity and a high COF. Besides, the wear scars become shallower and narrower with a decrease in $h$, indicating higher wear resistance.

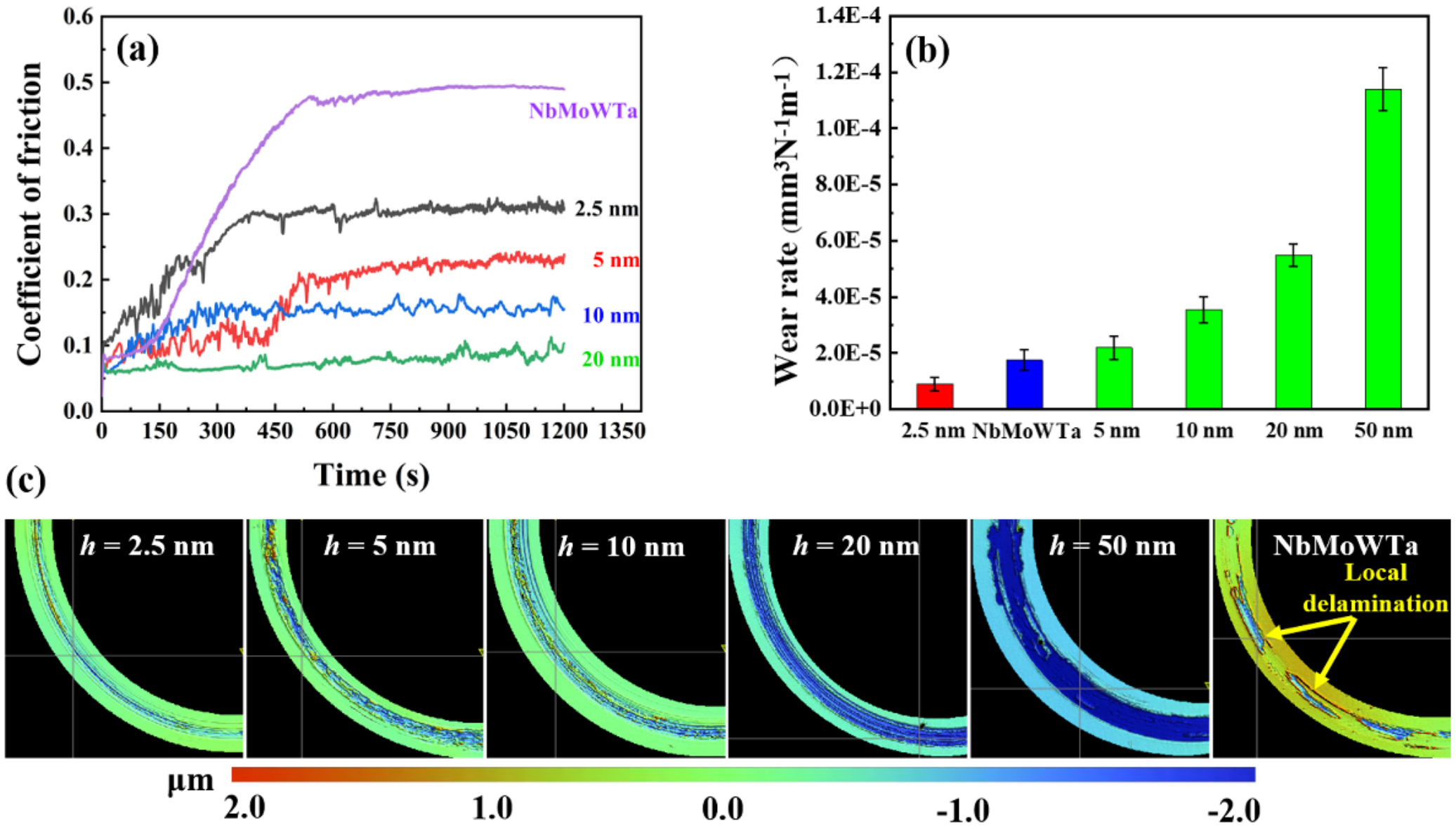

**Figure 8.** The tribological properties and surface profiles of all the NbMoWTa/Ag multilayers and monolithic NbMoWTa film. (a) COF and (b) wear rate of different films. (c) White light interferometry images of the worn surfaces.

In order to further explore the influence of Ag layer on the friction behavior, the SEM morphology of the wear track (Figure 9) after dry sliding tests is presented. The wear track of the NbMoWTa RHEA film shows avulsion, pulling-out and local fracture of the surface as shown in Figure 9a. In contrast, the micrographs of multilayers containing Ag are different and the degree of surface damage significantly decreases in NbMoWTa/Ag. During the gradual wear process, distinct patches of Ag-rich films appear on the worn surface, which can play an important role in improving the lubricating property[17,80]. Generally, these tribo-layers can effectively decrease the direct contact and shear strength between the film and the counter ball, and thus the adhesive wear of the material is inhibited[22,81]. On the other hand, it's obvious that with the change of $h$, the surface coverage of the lubricating film also changes. When $h$ is increased, a more integrated lubricating film is detected on the frictional surface and nearly full film coverage is achieved on the multilayer with $h$ = 20 nm (Figure 9e).

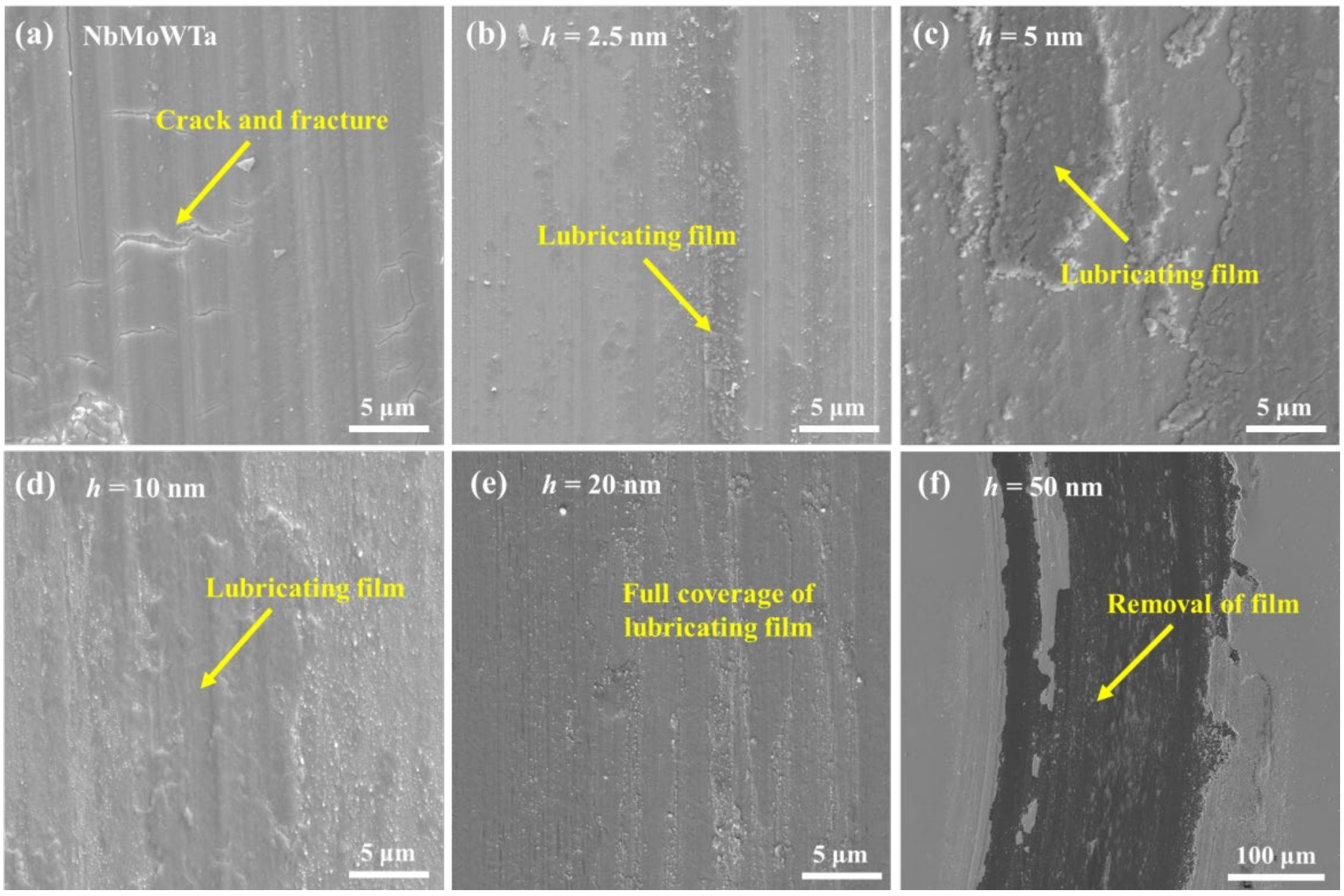

**Figure 9.** SEM micrographs of the worn surfaces of (a) monolithic NbMoWTa and NbMoWTa/Ag multilayers with *h* of (b) 2.5 nm, (c) 5 nm, (d) 10 nm, (e) 20 nm, (f) 50 nm.

XPS analysis of the lubricating films were further carried out to measure the chemical bonding of the worn surface from O 1s, Ag 3d, Nb 3d, Mo 3d, W 4f and Ta 4f peaks. In order to clarify the *h* dependent friction behavior, multilayers with *h* = 20 nm and 2.5 nm were chosen for direct comparison. In Figure 10a, the high-resolution XPS spectra show a strong oxide peak in addition to little hydroxide adsorbates. It is noted from Figure 10b that the Ag 3d spectra are deconvoluted into two overlapping peaks of metallic Ag and oxide compounds. The compounds are further identified as the composites of possible metal oxides and silver containing binary metal oxide phases ($Ag_xTM_yO_z$, where TM is the refractory transition metal) as shown in Figure 10c,e, respectively. Another effect of layer thickness is that it alters the relative content of silver oxides in the lubricating layer, as suggested by the area ratio of corresponding sub-peaks (Figure 10b). When *h* increases from 2.5 nm to 20 nm, the calculated ratio between the peak for Ag in its oxidation state over the peak for it in the metallic state increases, indicating an increase in the content of Ag oxides. These Ag oxides, especially the lubricious binary metal oxide formed by tribo-chemical reactions in the contact area, are helpful for reducing the COF[82]. Upon sliding, the weak Ag-O bond is easy to break and allows for interlayer shearing[83], thus contributing to the lubricating property and a smoother wear surface. On the other hand, these metal oxides are more easily formed on the surface provided that the supply of Ag is sufficient (e.g. in the multilayer with a larger thickness of Ag layer). A thicker Ag layer can provide more reactants and form a continuous lubricating film (Figure 9e), thus giving rise to the decreased COF.

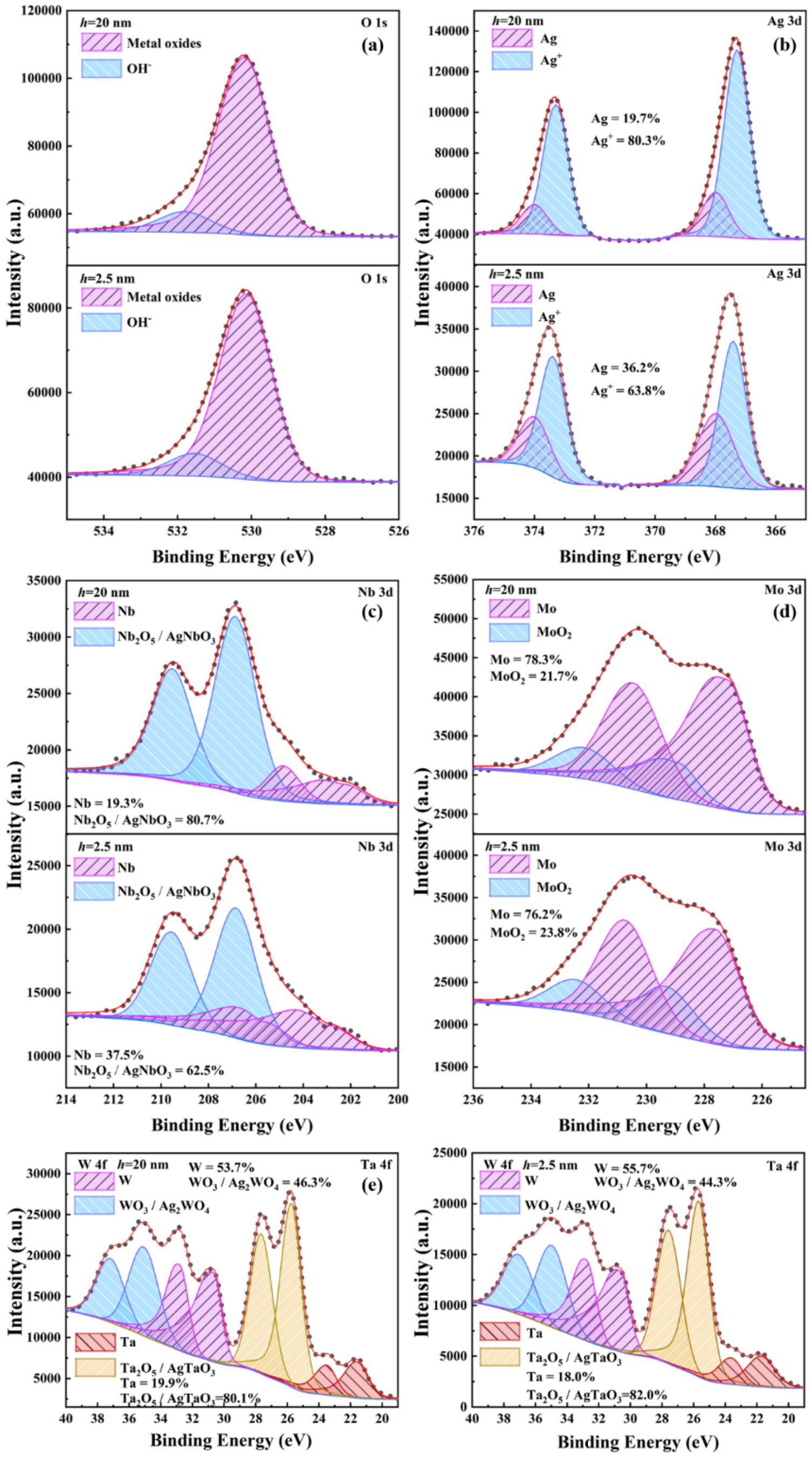

**Figure 10.** Deconvolution of high-resolution XPS spectrum of elements at the worn surfaces of NbMoWTa/Ag multilayers with $h$ = 20 nm and 2.5 nm, respectively. (a) O 1s, (b) Ag 3d, (c) Nb 3d, (d) Mo 3d and (e) Ta/W 4f.

Based on the above presented SEM and XPS results, it can be concluded that there are mainly two factors responsible for the layer thickness dependent friction behavior. First, adequate Ag can be supplied by the multilayer with a higher $h$ to form an integrated lubricating film. Second, the lubricity is also improved due to the formation of more oxides. Accordingly, for the multilayer with $h = 20$ nm, the best self-lubrication effect is achieved because full surface coverage of the lubricating film was obtained in the wear track. However, when $h$ is large, the mechanical properties of the multilayers degrade greatly, causing significant wear when compared with the monolithic NbMoWTa film possessing high hardness (Figure 8b). Therefore, just like the strength-ductility trade-off in alloys, the COF and wear rate also have a similar trade-off correlation in many self-lubricating composite films, as shown in Figure 11. The multilayered strategy in this work provides an optimal wear rate-COF combination, which surpasses most of the other Ag containing self-lubricating films (indicated by red arrow).

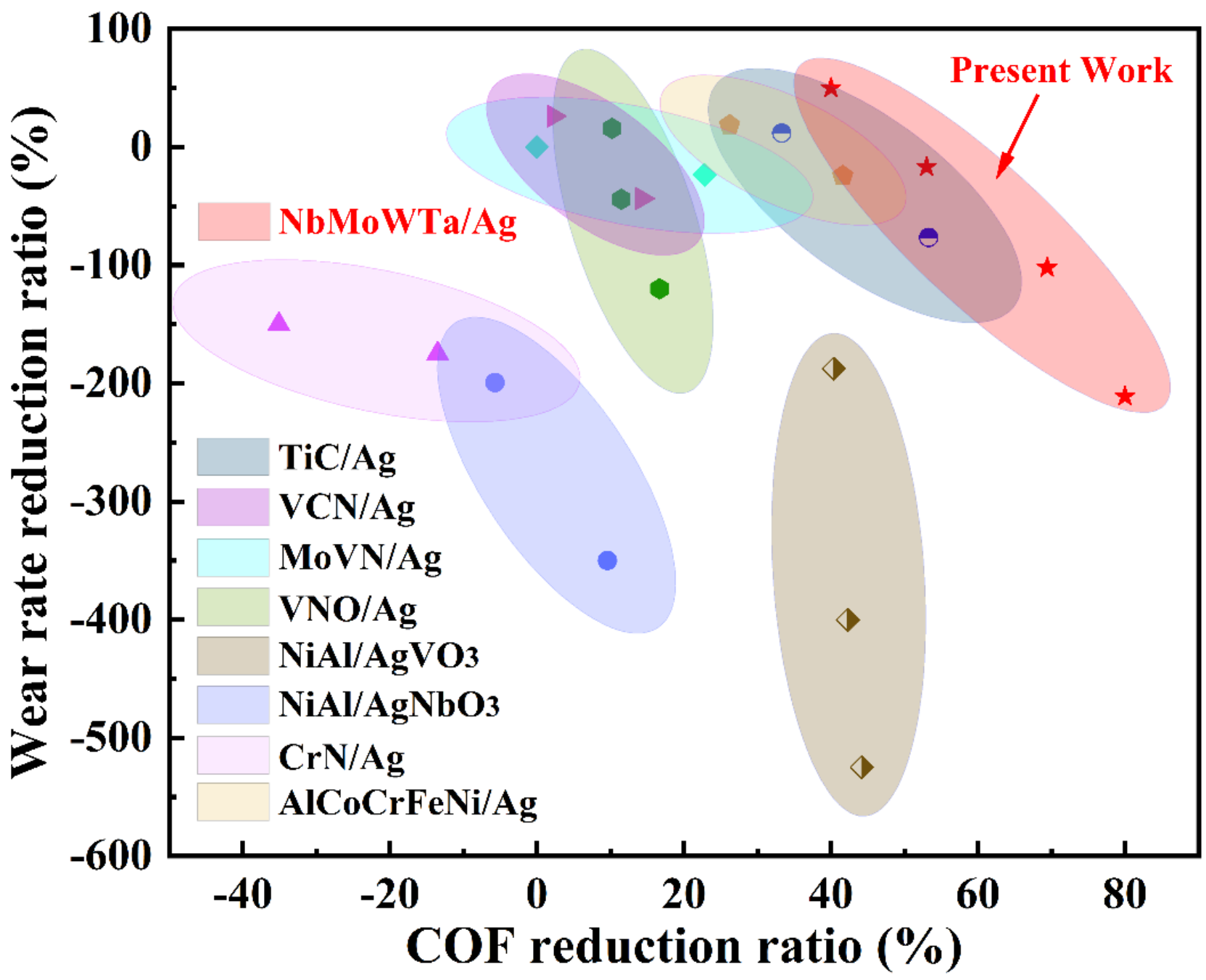

**Figure 11.** Wear rate reduction ratio of self-lubricating composite films reported in the literatures as a function of the COF reduction ratio. Data of these Ag contained self-lubricating films are chosen from Refs.[53-55,57,58,82,84,85]

Based on above results, the best tribological performance in the present work is obtained when the individual layer thickness $h$ is 2.5 nm. With a decrease of $H$ only by 14%, this multilayer shows a reduction of ~ 50% in the wear rate and ~ 40% in COF in comparison with the monolithic NbMoWTa film. The wear property of the NbMoWTa film is related to its hard but brittle characteristic, which leads to a local failure of the film under high friction (Figure 9a). In contrast, there is only slight adhesive and abrasive wear found for the multilayer with $h = 2.5$ nm (Figure 9b), due to the buildup of a lubricating layer along with a high load bearing ability of the surface. An effort was also made to prepare a multilayer with $h$ smaller than 2.5, but similar to those reported previously[31,52], significant intermixing between constituent layers occurred, leading to a broken of layered structure and a degradation of the mechanical properties. Therefore, the best wear performance of NbMoWTa/Ag multilayer can only be achieved when $h = 2.5$ nm.

Finally, it should also be noted that the counterbody ball can be protected from excess wear when sliding against the multilayer. The white light interferometry topographies of $Si_3N_4$ balls from the tribo-tests are given in Figure 12. The $Si_3N_4$ ball worn against NbMoWTa shows abrasive and adhesive wear with little material transfer. In contrast, obvious material transfer to the counterbody ball in the contact zone is observed for the multilayer with $h = 2.5$ nm, as outlined in Figure 12a. This is further confirmed by SEM and EDS micrographs as shown in Figure 13a. Ag is more prominent in the transfer layer than Si and N, compared to the unworn region. During sliding, the solid lubricant in the multilayer is easily adhered to the counter ball, thereby generating the lubricating film and reducing wear.

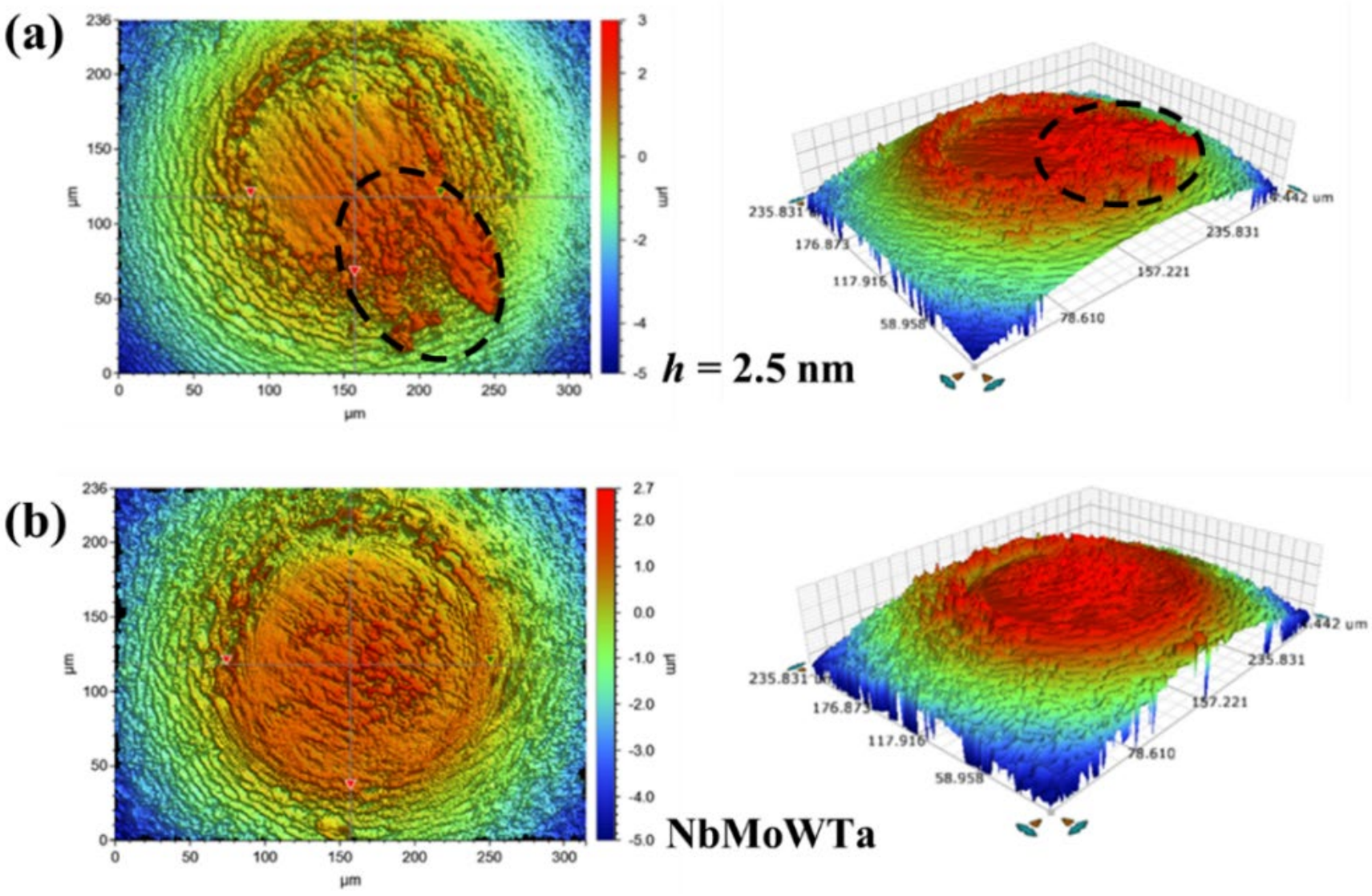

**Figure 12.** White light interferometry images of the worn surface of $Si_3N_4$ counterbody for (a) the multilayer with $h = 2.5$ nm and (b) the monolithic NbMoWTa film, respectively.

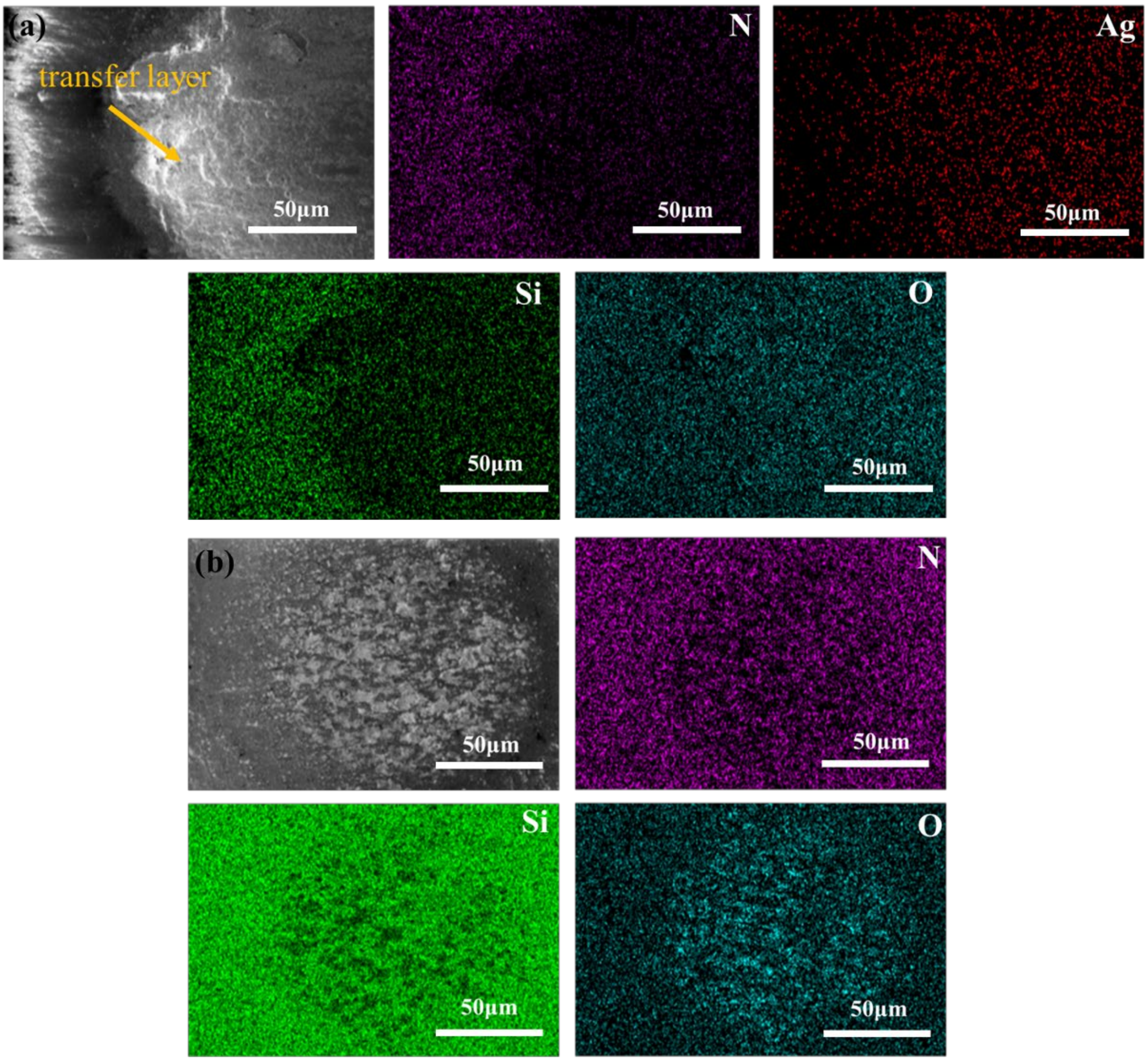

**Figure 13.** SEM images and element distribution maps of (a) the multilayer with $h$ = 2.5 nm and (b) the monolithic NbMoWTa film, respectively.

Summing up, the multilayer shows potential for the improvement of tribological properties relative to current RHEA-based films and may be suitable for use under severe frictional conditions because of the high hardness brought about by the multilayered structure. To fabricate this wear-resistant HEA, two prerequisites needed to be satisfied: (1) selecting a HEA with excellent mechanical properties; (2) introducing a solid lubricant with moderate lattice misfit at the interfaces ($<$ 5%); (3) the layer thickness of the solid lubricant should be modulated to achieve optimal wear performance.

## 4. CONCLUSIONS

In this work, a RHEA NbMoWTa/Ag composite film with a laminated structure has been designed and prepared. Detailed microstructural characterization, mechanical properties and tribological behavior of the multilayers with different individual layer thickness $h$ were systematically studied. As a result of structural transformation of sublayer, coherent interfaces were therefore obtained as $h$ decreases from 100 nm to 2.5 nm. Based on these structural transitions, a peak hardness of ~ 9.4 GPa was achieved in the multilayer, which is 52% higher than the prediction by the rule of mixture. The transition of the deformation mechanism from classic Hall–Petch strengthening to coherent strengthening was found to contribute to the monotonic increase in hardness for the NbMoWTa/Ag multilayers. Furthermore, the multilayer with $h$ = 2.5 nm shows a significant reduction in both the wear rate and the COF when compared with the monolithic NbMoWTa film. This can be attributed to the formation of a durable lubricating film and a high load bearing ability of the surface.

This study will motivate the design and fabrication of more novel RHEA films with excellent mechanical and tribological properties, which could be valuable in guiding the long-term performance of HEAs for industrial applications.

**ASSOCIATED CONTENT**

**Supporting information**

HAADF-STEM and corresponding EDS mapping of the NbMoWTa/Ag multilayer with $h$ = 100 nm and 5 nm; Typical cross-sectional TEM images of NbMoWTa/Ag multilayer with $h$ = 5 nm and 3D white light interferometry images of the worn surfaces of different films.

**Declaration of competing interest**

There are no conflicts to declare.

**ACKNOWLEDGMENTS**

The authors would like to thank the Natural Science Foundation of China (Nos. 51801161, 51975474), Shanghai Sailing Program and the Fundamental Research Funds for the Central Universities (3102019JC001).

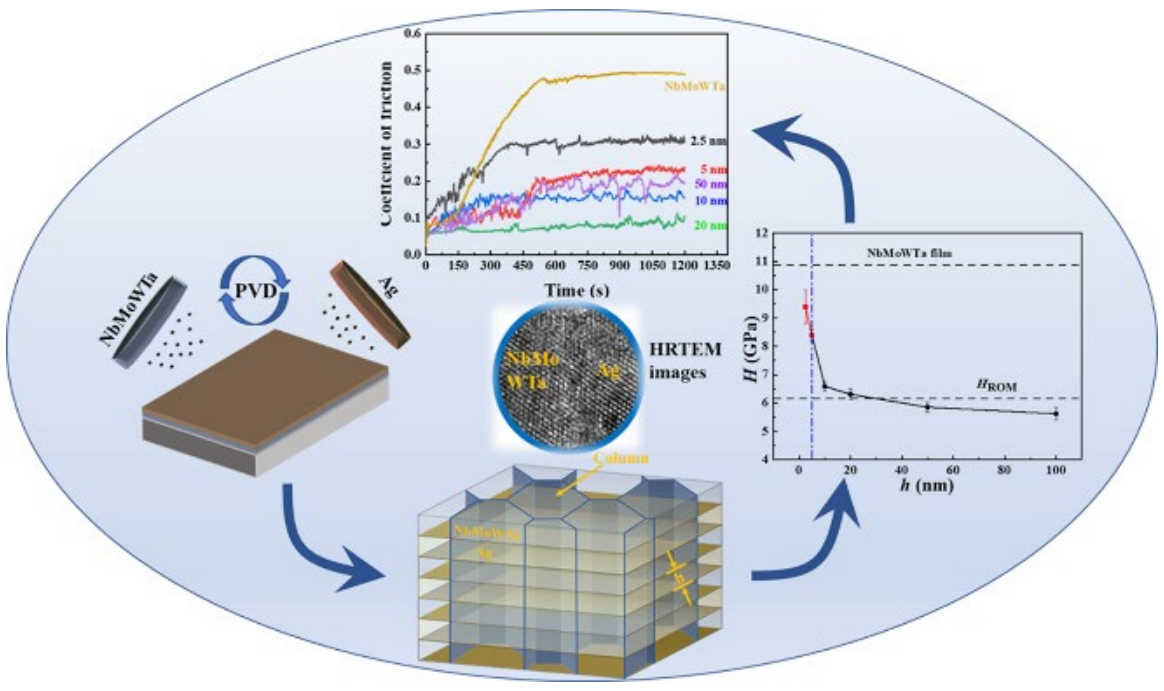

Abstract Graphics